\documentclass[twocolumn,pre,showpacs,showkeys,preprintnumbers,amsmath,amssymb]{revtex4}

\usepackage[lofdepth,lotdepth,caption=false]{subfig}
\usepackage{graphicx}
\usepackage{dcolumn}
\usepackage{bm}

\usepackage{booktabs}
\usepackage{amsmath}
\usepackage{epstopdf}

\def\Im{\mathop{{\cal I}\! m}}
\def\Re{\mathop{{\cal R}\! e}}

\def\ie{\hbox{\it i.e.,}{\ }}

\def\beq{\begin{equation}}
\def\eeq{\end{equation}}
\def\partder#1#2{{\partial #1\over\partial #2}}

\begin{document}

\preprint{DRAFT}

\title{Plasma Effect in The Longitudinal Space Charge Induced 
Microbunching Instability for Low Energy Electron Beams}

\author{Dazhang Huang}
\email{huangdazhang@sinap.ac.cn}
\author{Qiang Gu}%
\affiliation{%
Shanghai Institute of Applied Physics, Chinese Academy of Sciences, Shanghai, 201800, China P.R.\\
}%

\author{King Yuen Ng}
\email{ng@fnal.gov}
\affiliation{
Fermi National Accelerator Laboratory, Batavia, Illinois 60510, USA\\
}

\date{\today}

\begin{abstract}
The microbunching instability ($\mu$BI) usually exists in the LINAC of a free 
electron laser (FEL) facility. In many cases, the longitudinal space charge (LSC) is a dominant factor that 
generates the instability. For the highly bright electron beams, the plasma effect is found to be non-trivial in the development of the instability. In this paper, starting from the Vlasov and Poisson equations in the multiple-dimensional phase space, we perform the straightforward analysis of the microbunching instability based on the explicit formula of the longitudinal electric field introduced by the density perturbation in the longitudinal direction, in such a way to be highly comparable to the well-developed method for higher energy beams. This method generally applies in both the cases with and without acceleration and independent of lattice components. The results show that for a electron beam with small transverse emittance at low energies, which is always the case in the injector of a free electron laser device, the plasma effect results in the oscillation of the longitudinal electric field in the modified plasma frequency that depends on the transverse size of the beam, and the Landau damping effect in the longitudinal electric field due to the uncorrelated longitudinal velocity spread during the beam transportation. These two effects both play important roles in the development of the instability. As the result, the energy modulation driven by the LSC impedance differs from the regular value significantly and the discrepancy leads to the noticeable change of the final gain of the instability. 
\end{abstract}

\maketitle

\section{INTRODUCTION}
The possibility of oscillation in a plasma due to local separation of charges and the consequent restoring forces was discussed by J. D. Jackson long time ago.~\cite{JacksonJNE} The theory is based on a neutral plasma, which has both positively (ion) and negatively (electron) charged components. For a charged particle beam in an accelerator, although it is not neutral in terms of charges, there is still density fluctuation due to the graininess of the individual particles --- in our case, the individual electrons. Such graininess is usually smoothed out in the fluid model and ignored in most computations. In a highly intensive beam, however, it may introduce the ``plasma-like'' oscillation (for convenience, ``plasma oscillation'' is used hereafter). As of today, people have concluded that the plasma effect is not significant at high energies~\cite{ZHuang} and the physical model has already been implemented into the successful particle tracking codes such as {\footnotesize ELEGANT}.~\cite{elegant} Meanwhile, the work of great importance has been done for analyzing the plasma oscillation in a thermal electron beam in 3-D,~\cite{Mari1} and the full-dimensional analytical study of the microbunching instability developed in the space without acceleration for a longitudinally quasilaminar beam also reaches a very good agreement with the simulation results.~\cite{Mari2} In spite of that, a general method is still needed to connect the plasma effect to the gain of the instability no matter whether acceleration exists and independent of lattice components. Moreover, at low energies that the quasilaminar approximation does not hold very well, the discussions for the development process of the microbunching instability ($\mu$BI) are still not so adequate and need more efforts to reveal more detail. On the other hand, although a Particle-In-Cell (PIC) code like {\footnotesize IMPACT-T}~\cite{impact} is able to emulate the evolution of the beam accurately at low energies, an analytical way comparable to the well-developed method to compute the gain of the microbunching instability~\cite{Saldin} is also desired to perform a quick estimate for the instability.  

The discussion in this article is in a free space without any boundary. We start our discussion in the 6-D phase space by employing Vlasov and Poisson (Gauss) equations which describe the evolution of the distribution function of the electron bunch and the electric field induced by the charge distribution.  We then use a method similar to Jackson's~\cite{JacksonJNE} to linearize the Vlasov equation with 1-D approximation and obtain the solution of the initial-value problem. The solution includes the contributions from both the perturbed and unperturbed parts of the initial distribution, with the contribution from the velocity distribution included. In section~III, the electric field due to the density perturbation is derived by combining the solution of the initial-value problem and the Poisson equation~\cite{Venturini} together. In section~IV, we carry out the investigation by employing the Klimontovich distribution that describes the exact phase space location of each individual particle as the initial perturbation. We find that the evolution of the perturbed electric field is subject to periodic oscillation and Landau damping at low energies. In the last section, the damping factor as a function of modulation wavelength, beam energy and uncorrelated velocity spread for a longitudinally cold beam is illustrated, and the amplitudes of the energy modulation driven by the LSC impedance with and without the plasma effect included are computed for a typical linac lattice starting from the exit of the electron gun and ending before the compression. The results show that the difference between the energy modulations calculated by the different methods develops rapidly at low energies and becomes very large before the bunch compressor, and leads to the noticeable change to the gain of the instability. The energy modulation and the final gain are also computed as the functions of critical variables such as modulation wavelength and charge density. 
The summary and conclusion remarks are given at the end.  

\section{SOLUTION OF INITIAL VALUE PROBLEM}    
We carry out the investigation with the equations describing the evolution 
of beam distribution perturbed by the small amplitude space charge oscillation under the influence of the electromagnetic force. 
The discussion is in laboratory frame hereafter. Considering the derivations of the particle coordinates with respect to the reference particle due to the perturbation introduced by the oscillation, using the time evolution $t$ of the reference particle as the evolution nable, and assuming an cylindrically symmetric uniform beam, we focus our discussions on the phase space $(\Delta z, r, \Delta v_z, v_\perp)$ at time $t$ in a cylindrical coordinates system, where $\Delta z=z-z_0$ is the longitudinal coordinate of the target particle with respect to the reference particle, $r$ is the transverse coordinate of the target particle, $\Delta v_z=v_z-v_0$ is the longitudinal velocity of the target particle relative to the reference particle, and $v_\perp$ is the transverse velocity of the target particle. Here we also assume that the transverse location and the velocity of the reference particle are both zero, which is reasonable in the following investigation. Therefore in the phase space we are discussing, the linearized Vlasov-Poisson equation and the Poisson (Gauss) 
equation can be written:     
\beq
\frac{\partial f_1}{\partial t}+\Delta v_z\frac{\partial f_1}{\partial \Delta z}
+v_\perp\frac{\partial f_1}{\partial r}
-\frac{eE_z}{\gamma^3 m}\frac{\partial f_0}{\partial \Delta v_z}
+\frac{F_\perp}{\gamma m}\frac{\partial f_0}{\partial v_\perp}=0,
\label{Vlasov1}
\eeq
\beq
\Bigg[\frac{1}{r}\left(r\frac{\partial}{\partial r}\right)+\frac{1}{r^2}\frac{\partial^2}{\partial\phi^2}
+\frac{\partial^2}{\partial \Delta z^2}\Bigg]\Phi(r,\phi,\Delta z)=\frac{e\lambda}\epsilon
\rho(r,\phi,\Delta z),
\label{Poisson}
\eeq
where $-e$ is charge of an electron, $m$ is the electron mass, and the longitudinal electric field is given by
$E_z=-\partial\Phi/\partial\Delta z$. 
Because of the way Gauss law
is written, the normalization of the particle distribution is
\begin{align}
&\int \Big[f_0(\Delta v_z,\Delta z,x,y)+f_1(\Delta v_z,\Delta z,x,y,t)\Big] d\Delta v \nonumber \\
&~~~=\rho(\Delta z,x,y,t) \nonumber \\
&~~~=\rho_\perp(x,y)\rho_z(\Delta z,t),
\end{align}
where 
\beq
\int \rho_\perp dxdy=1
{\rm ~~and~~}
\int \rho_z(\Delta z)d\Delta z=L,
\eeq
with $L$ being the length of the electron bunch.
Note that $\rho(\Delta z,x,y,t)$ does not have the usual dimension of
$L^{-3}$; instead it has the dimension of $L^{-2}$.  
Here, $\rho_z(\Delta z)$ is dimensionless.  For a bunch with uniform longitudinal
distribution, $\rho_z(\Delta z)=1$.
By the same token we have
\begin{align}
&\int\lambda\Big[f_0(\Delta v_z,\Delta z,x,y,t)+f_1(\Delta v_z,\Delta z,x,y,t)\Big] \nonumber \\
&~~~\times d\Delta vd\Delta zdxdy \nonumber \\
&~~~=N_e,
\end{align}
the number of particles in the bunch.  We can also factorize the transverse
distribution as
\beq
f_0(\Delta v_z,\Delta z,x,y)=f_\perp(x,y)f_{z0}(\Delta v_z,\Delta z)
\eeq
and
\beq
f_1(\Delta v_z,\Delta z,x,y,t)=f_\perp(x,y)f_{z1}(\Delta v_z,\Delta z,t),
\eeq
where $f_\perp(x,y)=\rho_\perp(x,y)$ is the transverse distribution or
transverse density. For a transversely uniform beam, $f_\perp(x,y)=1/\pi r_b^2$, where $r_b$ is the beam radius. 

The Gauss's law or Poisson equation, Eq.~(\ref{Poisson}), 
will be solved by using the Green's function in the beam frame~\cite{Venturini} and transformed back to the laboratory frame in the next section. In the derivation of Eq.(\ref{Poisson}), we employ 
\beq
\frac{d\Delta v}{dt}=\frac{d\Delta\beta}{d\Delta z}\beta c^2{\rm~~~and~~~}\frac{d\Delta\beta}{d\Delta z}=-\frac{1}{\beta\gamma^3}\frac{eE_z}{mc^2}
\eeq  
where $\beta$ is the particle velocity divided by the speed of light and $\gamma=1/\sqrt{1-\beta^2}$ is the Lorentz factor.

Since the transverse velocity $v_\perp$ is small, we can assume
$\Delta v_z\approx \Delta v$ and $F_\perp\ll F_z$. Then Eq.~(\ref{Vlasov1}) 
simplifies to the one-dimensional form
\beq
\frac{\partial f_1}{\partial t}+\Delta v\frac{\partial f_1}
{\partial \Delta z}-\frac{eE_z}{\gamma^3 m}\frac{\partial f_0}{\partial\Delta v}=0.
\label{Vlasov2}
\eeq

Let us focus on Eqs.~(\ref{Vlasov2}) and (\ref{Poisson}). 
Following Jackson,~\cite{JacksonJNE} we perform Fourier transform in $\Delta z$ 
and Laplace transform in $t$ on Eq.~(\ref{Vlasov2}), and integrate by parts 
to obtain
\begin{align}\label{Jackson2}
\int\!\!d\Delta z\Bigg[&e^{-ik\Delta z+i\omega t}f_1(\Delta v,\Delta z,t)\Bigg]_{t=0}^{t=\infty} 
\!\!\! \nonumber \\
&+\!\int_{-\infty}^{\infty}\!\!d\Delta z\!\int_0^\infty\! e^{-ik\Delta z+i\omega t}dt 
\nonumber \\
&\times\Bigg[(-i\omega+ik\Delta v)f_1-\frac{e}{\gamma^3 m}
\frac{\partial f_0}{\partial \Delta v}E\Bigg]=0.
\end{align}
For $\omega$ in the upper half plane,  
the upper limit of the first term on the right hand side of 
Eq.~(\ref{Jackson2}) 
vanishes as $t\to\infty$. We have then the solution in $(\Delta v,\omega,k)$ space,
\begin{align}\label{f1}
\tilde f_1(\Delta v,\omega,k)&=\frac{1}{i(k\Delta v-\omega)}\Bigg[\tilde \Phi(\Delta v,k) \nonumber \\
&+\frac{e}{\gamma^3 m}\frac{\partial f_0}{\partial \Delta v}E(\omega,k)\Bigg],
\end{align} 
where
\beq
\tilde \Phi(\Delta v,k)=\int_{-\infty}^\infty d\Delta ze^{-ik\Delta z}f_1(\Delta v,\Delta z,t=0).
\label{phi}
\eeq
is the Fourier transform of the initial perturbation, or is called the bunching factor.~\cite{ZHuang}

Both Eq.~(\ref{f1}) and Eq.~(\ref{phi}) form the solution depending on 
the initial value of the density perturbation. If we perform inverse 
Fourier transform on $\omega$, we will obtain the density perturbation 
$f_1(\Delta v,t,k)$ at later time, which represents the time revolution of the 
density fluctuation. In the regular LSC theory, the density fluctuation 
is neglected. However, it will be taken into account under certain conditions
in the following discussions.

\section{ELECTRIC FIELD INDUCED BY LSC}
We solve Gauss law in the rest frame of the beam first and later
transform the result to the lab frame.  First, the Green's function for the
potential defined as the solution of~\cite{JacksonEM}
\begin{align}
&\left[\frac{1}{r}\left(r\partder{}{r}\right)+\frac{1}{r^2}\partder{^2}{\phi^2}
+\partder{^2}{z^2}\right]g(\vec r,\vec {r'}) \nonumber \\
&=-\frac{1}{r}\delta(r-r')\delta(\phi-\phi')\delta(z-z'),
\end{align}
is given by~\cite{Venturini, JacksonEM}
\beq
g(\vec r,\vec {r'})=\frac{1}{4\pi^2}\sum_m\int dk
e^{im(\phi-\phi')}e^{ik(z-z')}I_m(kr_<)K_m(kr_>).
\eeq
where $r_<$ and $r_>$ denotes the smaller and larger between $r$ 
and $r'$, respectively, and $I_m$ and $K_m$ are the modified Bessel 
function of the first and second kind.
Therefore,
\beq
\Phi(\vec r)=\int d^3\vec r g(\vec r,\vec {r'})
\left[\frac{-e\lambda\rho(\vec{r'})}{\epsilon_0}\right].
\eeq
and the longitudinal electric field reads 
\begin{align}
E_z(\vec r)&=\frac{e\lambda}{\epsilon_0}\int d^3\vec{r'}
\frac{i}{4\pi^2}\sum_m\int dk k \nonumber \\
&~~~\times\rho(\vec{r'})e^{im(\phi-\phi')}e^{ik(z-z')}I_m(kr_<)K_m(kr_>).
\end{align}
Make the assumption of the factorization of the transverse distribution.
We obtain
\begin{align}
E_z(\vec r)&=\frac{ie\lambda}{4\pi^2\epsilon_0}
\sum_m\int dk k\int r'dr'd\phi dz' \nonumber \\
&~~~\times f_\perp\rho(z')
e^{im(\phi-\phi')}e^{ik(z-z')}I_m(kr_<)K_m(kr_>).
\end{align}

We are interested in the longitudinal electric field at the beam axis.
Thus $I_m(kr_<)=1$ and $K_m(kr_>)=K_m(kr')$.
For uniform transverse distribution, $f_\perp$ is $\phi$-independent.
Thus $\phi$ can be integrated and only $m=0$ contributes, giving the result
\begin{align}
E_z(z)&=\frac{ie\lambda}{2\pi\epsilon_0}
\int dk k\int r'dr'dz'f_\perp\rho_z(z')
e^{ik(z-z')}K_0(kr')\nonumber\\
&=\frac{ie\lambda f_\perp}{2\pi\epsilon_0}
\int\frac{dk}{k}\big[1-\xi K_1(\xi)\big]\int dz'e^{ik(z-z')}\rho_z(z'),
\end{align}
where $\xi=kr_b$.  

Next transform to the lab frame. What we need to do is to
let $\xi=kr_b/\gamma$~\cite{Venturini} instead. Meanwhile, both $E_z$ and $\rho_z$ become
time dependent. Also instead of $z$, we use $\Delta z=z-z_0$, where $z_0$
is a reference position, for example, the bunch center. We arrive at
\begin{align}
E_z(\Delta z,t)&=-i\frac{e}{4\pi\epsilon_0}\frac{\lambda f_\perp}{\pi\gamma^2}\big[1-\xi K_1(\xi)\big]\int d\Delta v d\Delta z' \nonumber \\
&~~~\times\int dk k e^{ik(\Delta z-\Delta z')}f_1(\Delta v,\Delta z',t),
\end{align}

Now go to the $k$-$\omega$-space.  The Fourier transform gives,
\begin{align}
\tilde E_z(k,\omega)&=\int d\Delta z dt e^{-ik\Delta z+i\omega t}E_z(\Delta z,t)
\nonumber\\
&=\frac{ie\lambda f_\perp}{\epsilon_0k}\big[1-\xi K_1(\xi)\big]
\int d\Delta v \nonumber \\
&~~~\times\int dt e^{i\omega t}e^{-ik\Delta z'}f_1(\Delta v,\Delta z',t)
\nonumber\\
&=\frac{ie\lambda f_\perp}{\epsilon_0k}\big[1-\xi K_1(\xi)\big]
\int d\Delta v\tilde f_1(\Delta v,k,\omega).
\end{align}
Note that the unperturbed part $f_0(\Delta v,\Delta z')$ does not
contain the high-frequency modulation component and therefore does not
contribute in above.  The assumption is that the unperturbed distribution
$f_0(\Delta v,\Delta z')$ is {\em smooth}. 

The Fourier transformed perturbed distribution $\tilde f_1(\Delta v,k,\omega)$ 
from Eq.~(6) is now substituted to arrive
\begin{align}\label{Ez1}
\tilde E_z(k,\omega)&=\frac{e\lambda f_\perp}{\epsilon_0\epsilon_rk}
\big[1-\xi K_1(\xi)\big]\nonumber\\
&\times\int_W d\Delta v\int_{-\infty}^\infty d\Delta z
\frac{e^{-ik\Delta z}f_1(\Delta v,\Delta z,t=0)}{k\Delta v-\omega}, 
\end{align}
with the relative dielectric factor (permittivity) given by
\begin{equation}
\epsilon_r=1-\frac{e^2\lambda f_\perp}{\epsilon_0\gamma^3mk^2}
\big[1-\xi K_1(\xi)\big]
\int_W\frac{\partial g_0}{\partial\Delta v}\frac{d\Delta v}{\Delta v-\omega/k}.
\label{permittivity}
\end{equation}
The path of the integration $W$, depicted in Fig.~1, is from $\Delta v=-\infty$
to $\infty$, passing below the pole $\Delta v=\omega/k$. This path comes from the analytic continuity from the upper 
$\omega$-half-plane to the whole $\omega$-plane. Introducing the frequency of plasma oscillation 
in laboratory frame, $\omega_p=\sqrt{e^2\lambda f_\perp/\gamma^3\epsilon_0 m}=\sqrt{e^2n_0/\gamma^3\epsilon_0 m}$, and defining the modified plasma frequency 
\beq\label{omegapmd}
\bar\omega_p=\omega_p\big[1-\xi K_1(\xi)\big]^{1/2}
\eeq
Eq.~(\ref{permittivity}) can be written as
\beq
\epsilon_r=1-\frac{\bar\omega_p^2}{k^2}\int_W\frac{\partial g_0}{\partial \Delta v}
\frac{d\Delta v}{\Delta v-\omega/k}.
\label{permittivity1}
\eeq 

Equation~(\ref{permittivity1}) is called the dispersion relation, 
it is a function of the wavenumber $k$ of the density fluctuation. 
Equation~(\ref{Ez1}) is the expression of the longitudinal electric 
field induced by the LSC under the influence of density fluctuation 
(plasma oscillation). Apparently, it includes the contribution due to 
the velocity distribution of the beam. 

The factor $1-\xi K_1(\xi)$ in the definition of the modified plasma frequency represents the transverse dependence of the longitudinal electric field due to LSC. For the regular parameters used in beam physics, the number of it is in between 0.1 and 1, and goes to 1 very rapidly as the beam radius goes from zero to $\sim$1 mm. Therefore we can confidently conclude that the value of the modified plasma frequency and that of the regular plasma frequency are almost in the same order for the normal beam parameters. On the other hand, in the pure 1-D discussion where the approximation $r_b\to\infty$ is employed, the dispersion relation degenerates to the familiar form.~\cite{JacksonJNE, Kim} In the following discussion, we will see that the modified plasma frequency plays a very important role in the development of $\mu$BI.

In most of the cases, where the initial momenta and locations of the electrons 
are decoupled, the perturbation takes the form 
$f_1(\Delta v,\Delta z,t=0)=f_{v1}(\Delta v,t=0)f_{z1}(\Delta z,t=0)$. Equation~(\ref{Ez1}) becomes
\begin{align}\label{Ez2}
\tilde E_z(k,\omega)&=\frac{e\lambda f_\perp\big[1-\xi K_1(\xi)\big]}
{\epsilon_0\epsilon_rk}\int_Wd\Delta v\frac{f_{v1}(\Delta v,t=0)}{k\Delta v-\omega} 
\nonumber \\
&\times\int_{-\infty}^\infty{e^{-ik\Delta z}f_{z1}(\Delta z,t=0)d\Delta z}. 
\end{align}
Equation~(\ref{Ez2}) decouples the contributions from the beam density 
distribution and velocity/momentum distribution.  

\section{INFLUENCE ON MICROBUNCHING INSTABILITY}
In this section, we start our discussions on the effects 
of the modified LSC impedance in microbunching instability. According to Z. Huang and Saldin, et al., for a beam with Gaussian energy distribution and taking into account the compression, the gain in density modulation reads~\cite{ZHuang, Saldin}
\begin{align}\label{gain}
&G=\Bigg |\frac{b_f}{b_o}\Bigg | \nonumber \\
&\approx \frac{CI_0}{\gamma I_A}\Bigg |k_fR_{56}\int_0^L{ds\frac{4\pi Z(k_0;s)}{Z_0}}\Bigg |\exp\Bigg (-\frac{1}{2}C^2k_f^2R_{56}^2\sigma_\delta^2\Bigg ).
\end{align}
where $C$ is the compression factor of a bunch compressor (chicane), 
$R_{56}$ is the transport matrix element of the whole bunch compressor, $\sigma_\delta$ is the relaive uncorrelated energy spread and $b_f$ and $b_0$ are the final and the initial bunching factor, respectively. $Z(k_0;s)$ is the impedance per unit length at the modulation wavelength $k_0$, $I_0$ is the initial beam current without density modulation and $I_A=17$ kA is the Alfven current. In Eq.~(\ref{gain}), the amplitude of the energy modulation takes the form
\beq
\Delta\gamma=\frac{I_0\rho_{\rm i}}{I_A}\Bigg |\int_0^L{ds\frac{4\pi Z(k_0;s)}{Z_0}}\Bigg |,
\label{dgammaeq}
\eeq
where $\rho_{\rm i}$ is the relative amplitude of the density modulation.

The impedance per unit length of the longitudinal space charge is defined by
\beq
E_z(k)=-Z(k)I(k),
\label{EzImp}
\eeq
where $I(k)$ is the Fourier transform of the beam current, \ie 
$I(k)=e\beta c\lambda\rho_z(k)$. 

Our discussion will be mainly based upon Eq.~(\ref{gain}), Eq.~(\ref{dgammaeq}) and Eq.~(\ref{EzImp}). In the following, we will show that the amplitude of the energy modulation $\Delta\gamma$ performs plasma-like oscillation in time domain. As the result, the form of the final gain will be changed.

In the next, our discussions will be focused on the low and the high energy cases. When computing the electric field $E(k,t)$ based on equation~(\ref{Ez1}), one can see that there are poles, $\epsilon_r(k,\omega)=0$ and $\omega=k\Delta v_j^0$, enclosed by the path of integration over $\omega$. When the beam energy is low, following Jackson~\cite{JacksonJNE}, and let $\omega/k=x+iy$, where $y$ is small, we have the zero of $\epsilon_r$ occurs when~\cite{JacksonJNE}
\beq
\frac{k^2}{\bar\omega_p^2}=\mathop{\cal P}\int\frac{\partial g(\Delta v)}{\partial\Delta v}\frac{d\Delta v}{\Delta v -x},
\label{realomega}
\eeq
and
\beq
\Im(\omega)=-\frac{\pi k\partial g(x)/\partial x}{\mathop{\cal P}\int\frac{\partial^2g(\Delta v)/\partial\Delta v^2d\Delta v}{\Delta v -x}},
\label{imgomega}
\eeq
where $\mathop{\cal P}$ represents the principal value of the integral.

\begin{figure}[htb]
   \centering
   \includegraphics*[width=60mm]{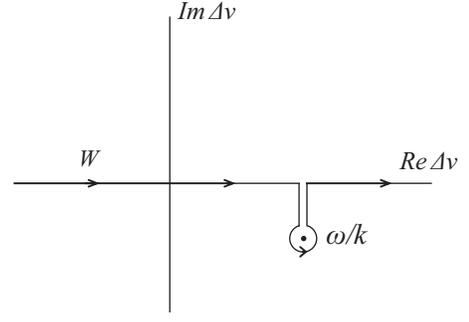}
   \vskip-0.1in
   \caption{The Landau contour $W$ for the definition of $\epsilon_r$, reproduced from reference~\cite{JacksonJNE}.}
   \label{LandauContour}
\end{figure}

Without losing generality, assuming the initial unperturbed velocity distribution of the electrons obeys Gaussian form
\beq
g_0(\Delta v)=\frac{1}{\sigma_v\sqrt{2\pi}}e^{-\Delta v^2/2\sigma_v^2},
\label{velocity}
\eeq
here we use $\sigma_v$ instead of $\sigma_{\Delta v}$ for convenience. Plugging Eq.~(\ref{velocity}) into the expression of relative permittivity defined in Eq.~(\ref{permittivity}), and first let us consider the real part of modulation frequency $\omega$, given that the imaginary part $\Im(\omega)$ is small, based on Eq.~(\ref{realomega}), we integrate over the Landau contour (Fig.~\ref{LandauContour})~\cite{JacksonJNE}, and employ the first order approximation, we obtain
\begin{align}
0&\approx1+\frac{\bar\omega_p^2}{\sqrt{2\pi}k^2\sigma_v^3}\Bigg (\int_W{e^{-\frac{\Delta v^2}{2\sigma_v^2}}d\Delta v} \nonumber \\
&\ \ \ \ \ \ +\frac{\Re(\omega)}{k}\int_W{e^{-\frac{\Delta v^2}{2\sigma_v^2}}\frac{d\Delta v}{\Delta v-\Re(\omega)/k}}\Bigg ) \nonumber \\
&=1+\frac{\bar\omega_p^2}{k^2\sigma_v^2}\Bigg [1+i\sqrt{\frac{\pi}{2}}\Bigg (\frac{\Re(\omega)}{k\sigma_v}\Bigg )w\Bigg (\frac{\Re(\omega)}{\sqrt{2}k\sigma_v}\Bigg )\Bigg ] \nonumber \\
\label{permittivity2}
\end{align}
where $w(z)$ is the complex error function.

In the low energy limit, $\omega/k\sigma_v\gg 1$, and $\bar\omega_p\gg k\Delta v_j^0$ for most of the $N_e$ particles. In this case, the complex error function can be expanded as
\beq\label{errorfunc}
w(z)=\frac{i}{\sqrt{\pi}z}\bigg [1+\frac{1}{2z^2}+\cdots \bigg ];
\eeq
therefore Equation~(\ref{permittivity2})) becomes
\begin{align}
0&\approx1+\frac{\bar\omega_p^2}{k^2\sigma_v^2}\Bigg [1+i\sqrt{\frac{\pi}{2}}\frac{\Re(\omega)}{k\sigma_v}\Bigg(\frac{i\sqrt{2}k\sigma_v}{\sqrt{\pi}\Re(\omega)}\Bigg) \nonumber \\ &\times\Bigg(1+\frac{k^2\sigma_v^2}{(Re(\omega))^2}+\cdots\Bigg)\Bigg] \nonumber \\
&=1-\frac{\bar\omega_p^2}{(\Re(\omega))^2}+{\rm higher~order~small~terms} \nonumber \\
&\approx 1-\frac{\bar\omega_p^2}{(Re(\omega))^2}.
\label{permittivity3}
\end{align}
Thus we have $\Re(\omega)=\pm\bar\omega_p$.

On the other hand, according to Landau and Jackson, for a single-humped distribution of $g(\Delta v)$ like Gaussian, provided the damping is small, we can derive~\cite{JacksonJNE, Landau}
\begin{align}\label{imgomega1}
\Im(\omega)&\simeq-\sqrt{\frac{\pi}{8}}\bar\omega_p\Bigg(\frac{\bar\omega_p}{k\sigma_v}\Bigg)^3\exp\Bigg(\frac{-\bar\omega_p^2}{2k^2\sigma_v^2}\Bigg) \nonumber \\
&\equiv-\eta,
\end{align}
where $\sigma_v=\sqrt{kT/m_e}$ is defined as the rms thermal velocity of electrons in one dimension,~\cite{JacksonJNE} and in our discussion, $\sigma_v$ is the local uncorrelated rms velocity spread in longitudinal.  $\eta$ here is introduced as the Landau damping rate. 

Therefore we conclude that the zero of $\epsilon_r$ occurs when $\omega=\pm\bar\omega_p-\eta i$. To solve the problem, we introduce the Klimontovich particle distribution at $t=0$ as the initial perturbation
\beq\label{Klimontovich}
f_1(\Delta v,\Delta z,t=0)=\frac{1}{\lambda}\sum_{j=1}^{N_e}{\delta(\Delta v-\Delta v_j^0)\delta(\Delta z-\Delta z_j^0)},
\eeq  
where $\Delta v_j^0$ and $\Delta z_j^0$ are the initial velocity and relative longitudinal position of the $j^{\rm th}$ particle. Applying Eq.~(\ref{permittivity3}) and Eq.~(\ref{imgomega1}) into Eq.~(\ref{Ez1}), substituting the Klimontovich noise distribution (Eq.(\ref{Klimontovich})), and carrying out the integral by employing the residual principle, we have the electric field in $(k,t)$ phase space  
\begin{align}
E_z(k,t)&= \frac{e}{2\pi\epsilon_0\pi kr_b^2}\big[1-\xi K_1(\xi)\big]\times \nonumber \\ 
&\int_Cd\omega\frac{e^{-i\omega t}}{\epsilon_r(k,\omega)}\sum_{j=1}^{N_e}\frac{e^{-ik\Delta z_j^0}}{k\Delta v_j^0-\omega}
\nonumber \\
&\approx\frac{ie}{\epsilon_0\pi kr_b^2}\big[1-\xi K_1(\xi)\big]\times \nonumber \\ 
&\Bigg[\sum_{j=1}^{N_e}\frac{e^{-ik\Delta z_j^0}\bar\omega_p}{2}\left(\frac{e^{-i\bar\omega_p t}}{\bar\omega_p-k\Delta v_j^0}+\frac{e^{i\bar\omega_pt}}{\bar\omega_p+k\Delta v_j^0}\right)e^{-\eta t} \nonumber \\ &+\sum_{j=1}^{N_e}\frac{e^{-ik(\Delta z_j^0+\Delta v_j^0t)}}{\epsilon_r(k,k\Delta v_j^0)}\Bigg].  
\label{Ez31} 
\end{align}
where $f_\perp=1/\pi r_b^2$ is applied.

The numerator of the second term in Eq.~(\ref{Ez31}) describes the phase space revolution of a beam particle, and the denominator represents the shielding effect introduced by the other electrons.~\cite{Kim} Based on the work done by Kim and Lindberg,~\cite{Kim} in the low energy limit the second term in Eq.~(\ref{Ez31}) is small and can be neglected. Therefore we have Eq.~(\ref{Ez31}) take the form
\begin{align}
E_z(k,t)&\approx\frac{ie}{\epsilon_0\pi r_b^2k}\bigg[1-\xi K_1(\xi)\bigg]\Bigg [\cos(\bar\omega_pt)\sum_{j=1}^{N_e}e^{-ik\Delta z_j^0} \nonumber \\
&+i\sin(\bar\omega_pt)\sum_{j=1}^{N_e}\frac{k\Delta v_j^0}{\bar\omega_p}e^{-ik\Delta z_j^0}\Bigg ]e^{-\eta t} \nonumber \\
&\approx\frac{ie}{\epsilon_0\pi r_b^2k}\bigg[1-\xi K_1(\xi)\bigg]\cos(\bar\omega_pt)e^{-\eta t}\sum_{j=1}^{N_e}e^{-ik\Delta z_j^0}. 
\label{Ez32}
\end{align}
Note that in the derivation of Eq.~(\ref{Ez32}), $\bar\omega_p\gg k\Delta v_j^0$ is applied, where in the first step the term of $\mathcal{O}((k\Delta v_j^0/\bar\omega_p)^2)$ is neglected; and in the second step, the term of $\mathcal{O}(k\Delta v_j^0/\bar\omega_p)$ is ignored.  

According to Eq.~(\ref{EzImp}) and applying the well-known formula of the LSC impedance~\cite{ZHuang, Venturini} with the relativistic $\beta$ taken into account, we have
\beq
I_{\rm pert}(k,t)=e\beta c\cos(\bar\omega_pt)e^{-\eta t}\sum_{j=1}^{N_e}e^{-ik\Delta z_j^0}. 
\label{Ik}
\eeq

Equation~(\ref{Ik}) tells us that the current perturbation oscillates for the time being with the modified frequency $\bar\omega_p$ which is a function of beam radius $r_b$, modulation wavenumber $k$ and beam energy $\gamma$. Meanwhile, it is also damped out gradually with the Landau damping rate $\eta$ at low energies. As the result, the accumulated energy modulation (Eq.~(\ref{dgammaeq})) induced by the LSC impedance deviates from the value obtained from the regular method, thus the integrated energy modulation \ie $\Delta\gamma$ will differ from the regular value, which changes the final gain at the linac exit. We will show the example in the following section.

As the energy becomes larger, $\Im(\omega)\to 0$. Taking the limit of $\omega/k\sigma_v\ll 1$, when the argument is small, the complex error function can be expanded as
\beq\label{errorfunc1}
w(z)=\sum_{n=0}^{\infty}\frac{(iz)^n}{\Gamma(1+n/2)}=1+\frac{2iz}{\sqrt{\pi}}+\cdots
\eeq
Thus
\beq\label{permittivity4}
\epsilon_r=1+\frac{\bar\omega_p^2}{k^2\sigma_v^2}\Bigg [1+i\sqrt{\frac{\pi}{2}}\Bigg(\frac{\omega}{k\sigma_v}\Bigg)-\Bigg(\frac{\omega}{k\sigma_v}\Bigg)^2+\cdots\Bigg].
\eeq
At $\omega\sim\bar\omega_p$,
\beq\label{permittivity5}
\epsilon_r=1+\frac{1}{k^2\bar\lambda_D^2}\Bigg(1+i\sqrt{\frac{\pi}{2}}\frac{1}{k\bar\lambda_D}-\frac{1}{k^2\bar\lambda_D^2}+\cdots\Bigg),
\eeq
where the modified Debye length is defined
\beq\label{Debyem}
\bar\lambda_D=\sigma_v/\bar\omega_p.
\eeq
Therefore we conclude that when the electron energy is high, $\epsilon_r(\omega)=0$ has no solution and $\epsilon_r$ is almost independent of $\omega$. Following the same way as the low energy case to integrate Eq.~(\ref{Ez1}) over $\omega$ and $\Delta v$, finally we have
\beq\label{Ez4}
E_z(k)\approx\frac{ie}{\epsilon_0\epsilon_r\pi r_b^2k}\big[1-\xi K_1(\xi)\big]\sum_{j=1}^{N_e}e^{-ik (\Delta z_j^0+\Delta v_j^0t)}.
\eeq 

Note that In the high energy limit $k\bar\lambda_D\gg 1$ and therefore $\epsilon_r\to 1$. As the result, we can see that at high energies, the electric field approximately takes the same form as the one without plasma oscillation. It is not out of surprise because as we have already known, the plasma frequency decreases as the energy grows, and the wavelength of the plasma oscillation behaves in the opposite way; as the energy goes larger, the wavelength becomes larger as well, eventually when the wavelength is much longer than the scale of the whole accelerator lattice, the plasma oscillation becomes to play a very little role and can be ignored. Therefore our discussion will be mainly focused in the low energy regime. 

\section{Example}
\begin{figure}[htb]
   \centering
   \includegraphics*[width=80mm]{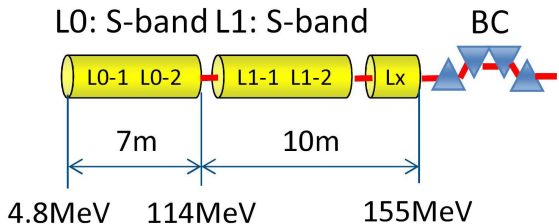}
   \vskip-0.1in
   \caption{(Color)The layout of the structure in discussion.}
   \label{layout}
\end{figure}

Let us now focus on Eq.~(\ref{dgammaeq}) and Eq.~(\ref{Ik}). Substituting $I_{\rm pert}$ in Eq.~(\ref{Ik}) into $I_0$ in Eq.~(\ref{dgammaeq}), one can see that the amplitude of the energy modulation oscillates in time with the modified plasma frequency $\bar\omega_p$ and damps with the damping factor $\eta$. Because both $\bar\omega_p$ and $\eta$ are functions of the beam energy, the numerical integration is needed to have the accumulated energy modulation. To demonstrate the problem, without losing generality, a typical structure that commonly appears in a FEL linac is employed with several acceleration sections including four S-band accelerating tubes for accelerating and energy chirping, one X-band structure for energy linearization and a few drift spaces in between. In order to be compatible with the analytical model derived above, we start our discussion at about 1.2 meters from the cathode, where the transverse size of the beam becomes much more stable than that when the beam is inside the gun. In the discussion, the beam is emitted from the cathode with the initial modulation amplitude of 10\%, when the beam comes out of the electron gun, the beam is longitudinally quasi-Gaussian and transversely uniform. To be consistent with the formulae developed in the last section, our dicussion is focused on the middle slice hereafter, which is approximately uniform longitudinally. The parameters are shown in Table~\ref{parameters}. The layout of the structure in our discussion is illustrated in Fig.~\ref{layout}.

\begin{table}
\caption{\label{parameters}beam parameters used in numerical integration taken from SXFEL~\cite{SXFEL}.}
\begin{ruledtabular}
\begin{tabular}{lllll}
 Parameter & Value \\
\hline
total beam charge (nC) & 0.5\\
charge in middle slice (nC) & 0.07 \\
initial beam energy ($\gamma$) & 10.5 \\
uncorrelated longitudinal velocity spread (m/s) & 285.5  \\
initial radius of middle slice (mm) & 0.485  \\
length of middle slice (mm) & 0.00026 \\
peak current (A) & 55 \\
relative slice energy spread before BC (\%) & 0.337 \\
beam energy before compression ($\gamma$) & 305 \\
modulation amplitude (\%)& 10 \\
$R_{56}$ of bunch compressor (mm) & -56.5 \\
compression ratio & 3 \\
\end{tabular}
\end{ruledtabular}
\end{table}     

\begin{figure}[htb]
   \centering
   \includegraphics*[width=60mm]{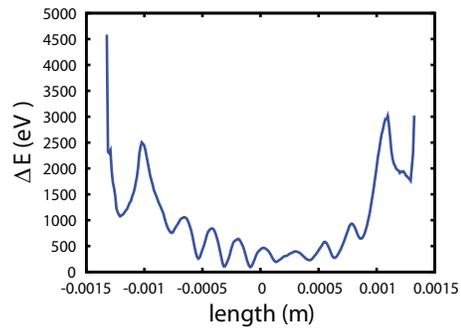}
   \vskip-0.1in
   \caption{The uncorrelated energy spread along the beam at the beginning of the computation by {\footnotesize IMPACT-T}, with FWHM $\approx 8$ ps.}
   \label{slice}
\end{figure}

Figure~\ref{slice} shows the uncorrelated energy spread ($\sigma_E$) along the beam in the beginning. We see that in the middle of the beam, $\bar{\sigma}_E\approx 550$ eV, which corresponds to $\sigma_v\approx 285.5$ m/s. According to reference~\cite{Mari2}, the longitudinal damping effect can be neglected when the electron displacement due to thermal motion ($d$) in a plasma period is much smaller than the longitudinal modulation wavelength ($\lambda$). In our case, based on the parameters in Table~\ref{parameters}, $d=2\pi\sigma_v/\omega_p\approx 7.7\times 10^{-7}$ m, whereas the typical number of the modulation wavelength in the microbunching problem $\lambda\sim1-10\times10^{-6}$ m, which is one order larger than $d$, therefore reference~\cite{Mari2} tells us that the longitudinal damping effect is not important. However, our computations show us that at short wavelengths, the damping effect is still non-trivial. 

\begin{figure}[htb]
\centering
\subfloat[(Color) Damping factor as a function of modulation wavelength and beam energy when $\sigma_v$ = 285 m/s.]{
   \includegraphics[width=0.7\linewidth]{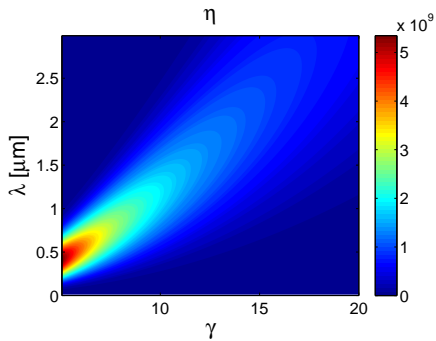}
   \label{subfig:fig1}
}

\subfloat[(Color) Damping factor as a function of modulation wavelength and local velocity spread when $\gamma=10.5$. ]{
   \includegraphics[width=0.7\linewidth]{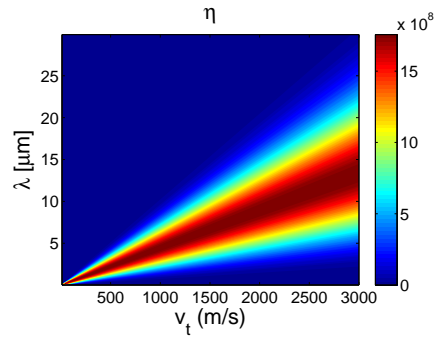}
   \label{subfig:fig3}
}
\caption[short for lof]{(Color) Landau Damping factor.}
\label{eta}
\end{figure}

The damping factor $\eta$ as a function of the modulation wavelength and the beam energy is illustrated in Fig.~\ref{eta}\subref{subfig:fig1} based on the parameters in Table~\ref{parameters}, one can see that it is very large at low energies and short wavelengths, and falls dramatically when energy goes higher and wavelength becomes longer. Thus the longitudinal Landau damping effect only play a role at low energies and vanishes rapidly at high energies. Fig.~\ref{eta}\subref{subfig:fig3} illustrates the damping factor as a function of the velocity spread at various wavelengths. We can see that as $\sigma_v$ goes high, the maximal of $\eta$ shifts to the longer wavelength. And in the most of the regions that concerns us, the damping factor is small enough to be ignored.

\begin{figure}[htb]
\centering
\subfloat[(Color) Energy modulation envelope ($\Delta\gamma$) with (green) and without plasma effect (blue) as a function of distance, at $\lambda=14\mu$m.]{
   \includegraphics[width=0.7\linewidth]{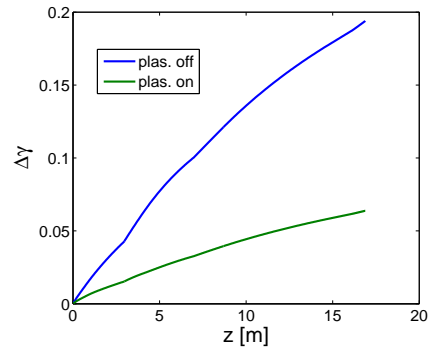}
   \label{subfig:fig1}
}

\subfloat[(Color) Energy modulation envelope as a function of distance and modulation wavelength. ]{
   \includegraphics[width=0.7\linewidth]{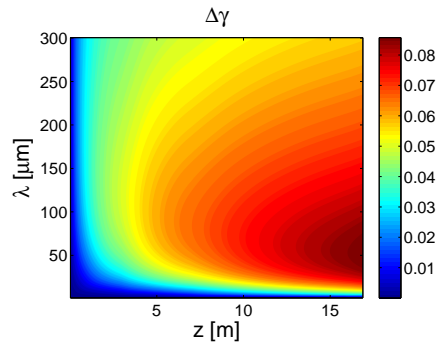}
   \label{subfig:fig3}
}
\caption[short for lof]{(Color) Evolution of the energy modulation envelope.}
\label{dgamma}
\end{figure}

As mentioned before, our discussion is for the middle slice of the beam which can be approximately accepted as an uniform cylinder. Fig.~\ref{dgamma} shows the evolution of the energy modulation envelopes with and without plasma oscillation. Figure~\ref{dgamma}\subref{subfig:fig1} shows the general behavior of the energy modulation at $\lambda=14\mu$m, and figure~\ref{dgamma}\subref{subfig:fig3} is the contour plot of the modulation as a function of the distance and the wavelength.  

\begin{figure}[htb]
   \centering
   \includegraphics*[width=60mm]{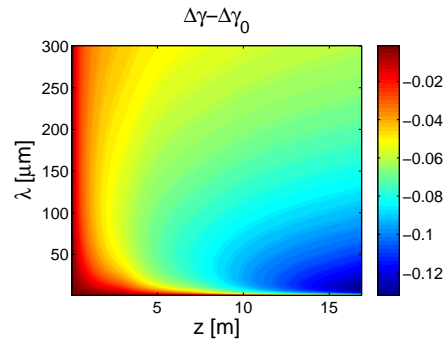}
   \vskip-0.1in
   \caption{(Color)The difference between the energy modulation envelopes with and without plasma effect as a function of distance and modulation wavelength .}
   \label{2ddgamma}
\end{figure}

The important thing that concerns us is the difference between the envelopes of the energy modulation with and without the plasma effect. Fig.~\ref{2ddgamma} gives us the difference at various modulation wavelengths and distances. In the figure we can see that the difference develops rapidly as the beam is being accelerated, which is consistent with our expection.   

\begin{figure}[htb]
   \centering
   \includegraphics*[width=60mm]{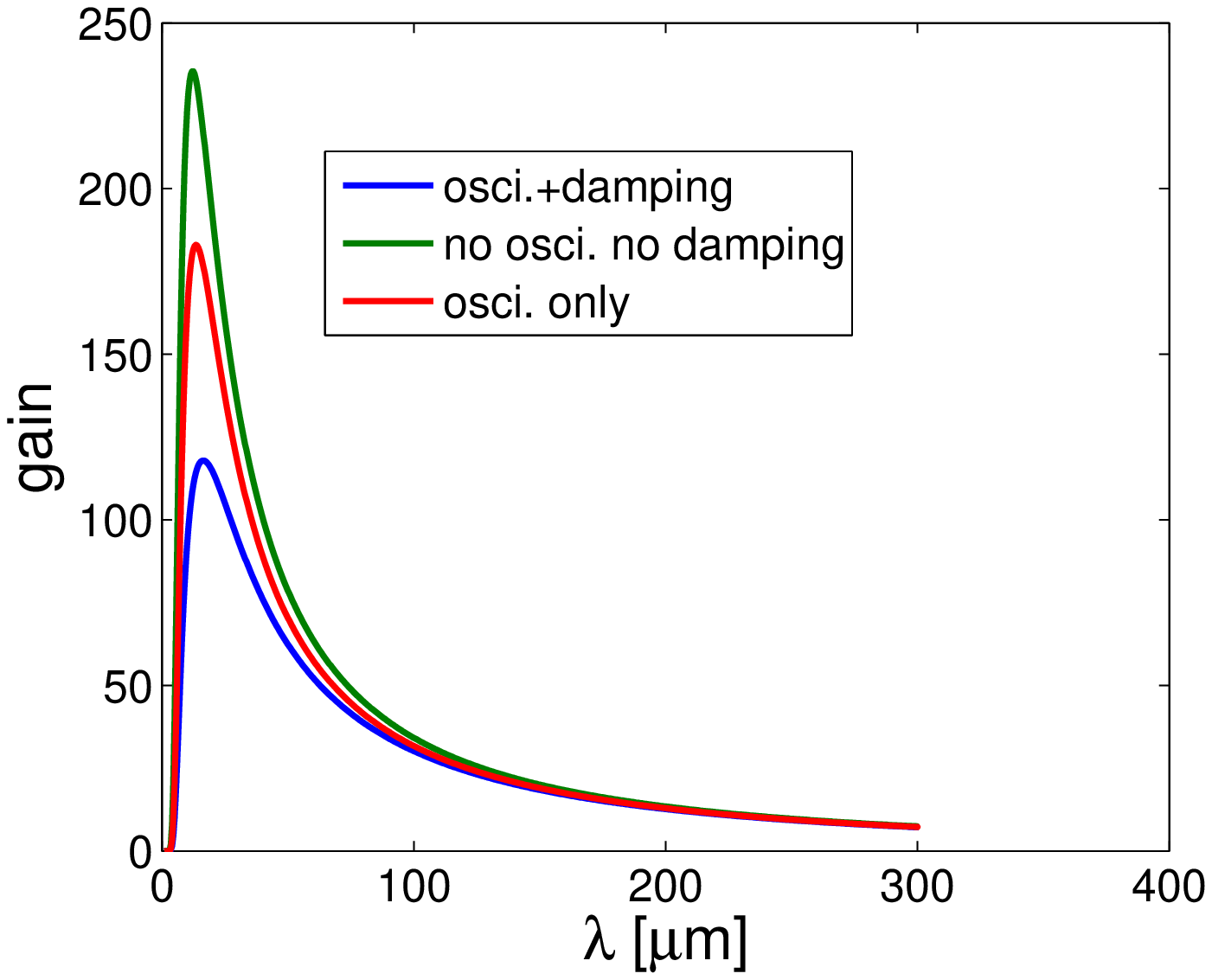}
   \vskip-0.1in
   \caption{(Color)The log plot of the  gain curves obtained at the exit with both plasma oscillation and damping (blue), with oscillation only (red), without any plasma effect (green).}
   \label{dgain}
\end{figure}

The difference between the amplitudes of the energy modulation leads to the discrepancy in the final gain. Figure~\ref{dgain} illustrates the gain curves (relative to the initial modulation amplitude)  obtained including the Landau damping and/or plasma oscillation, and with non of them included based on the parameters listed in Table~\ref{parameters}. In the figure, one can see that with the plasma effect included, the gain reduces very much. Also we can see that the Landau damping effect is much stronger at short wavelengths. 

\begin{figure}[htb]
\centering
\subfloat[(Color) Final gain pattern with plasma effect included as a function of beam density and wavelength.]{
   \includegraphics[width=0.7\linewidth]{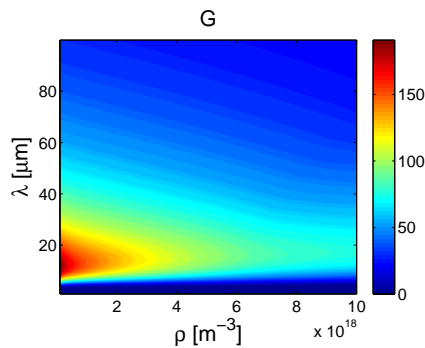}
   \label{subfig:fig1}
}

\subfloat[(Color) Discrepancy between the final gains with and without the plasma effect as a function of beam density and wavelength. ]{
   \includegraphics[width=0.7\linewidth]{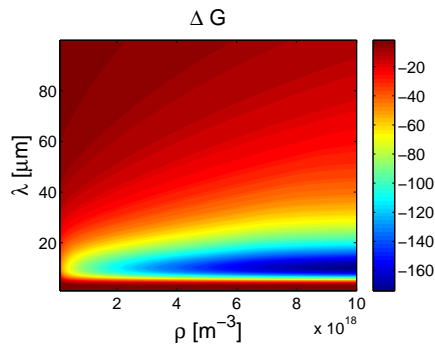}
   \label{subfig:fig3}
}
\caption[short for lof]{(Color) Final gain pattern (a) and its difference to the gain without plasma effect (b).}
\label{2dgain}
\end{figure}

Figure~\ref{2dgain}\subref{subfig:fig1} shows the pattern of the final gain as a function of the beam density and the wavelength. One can see that as the beam density goes smaller, the gain becomes larger. This is because the Landau damping effect in plasma decreases when the beam density is low. Meanwhile, figure~\ref{2dgain}\subref{subfig:fig3} tells us that the discrepancy between the gains with and without the plasma effect increases as the density rises and is very sensitive to it. This is not out of surprise because when the density is larger, the plasma effect becomes stronger and so does the discrepancy. 

\section{CONCLUSIONS}
In this paper, we investigated the plasma effect in the LSC-induced microbunching instability in an electron linac for an electron beam at low energies by analyzing the explicit expression of the longitudinal electric field introduced by the density perturbation in longitudinal. The electric field is derived by solving the Vlasov and Poisson equations, and degenerates to the classic solution when the beam energy becomes high. Our study shows that such an effect changes the gain of the instability. The general process of the instability starts from an electron beam with initial density modulation. As the beam propagates, the LSC impedance turns the density modulation into the energy modulation inside the beam, which is one of the most important factors to induce the instability in a significant amount. In our study, we find that the amplitude of the energy modulation performs the oscillation at the modified plasma frequency, which is usually not included in the gain computation. This plasma-like oscillation occurs mostly at low energies and brings noticeable change to the gain of the instability at short wavelengths. Meanwhile, the Landau damping of the longitudinal electric field during the beam transportation introduced by the uncorrelated longitudinal velocity spread also plays an important role in the development of the instability when the beam energy is low and the modulation wavelength is short, and starts to be trivial when the energy becomes higher and the wavelength goes long. The problem is demonstrated by building an example of an electron beam transported in a typical linac lattice, with the longitudinal phase space perturbation following Klimontovich form, and with all the parameters commonly used in free-electron-laser (FEL) physics. In the example, we find that the difference between the energy modulation amplitude with and without the plasma effect included develops rapidly as the beam is being accelerated, and brings a large discrepancy between the gain curve computed by the analytical model developed in this paper and the one by the regular way at short wavelengths.

At very high energies, our analysis shows that the plasma effect is rather trivial, which is consistent with the regular theory. However, the overall effect is non-trivial because the influence of the plasma effect developed at low energies exists all the way to the end. 

At last, it should be pointed out that there is another way to solve the problem. It begins with to solve for the growth directly from Vlasov equation and Poisson equation, because the imaginary part of the eign-frequency will give the growth rate. What is involved in the regular growth formula~\cite{ZHuang, Saldin}, for example the chirp and the compression, should be included in an equation of motion to be substituted into the Vlasov equation. In fact, without compression and chirp, the plasma frequency will be real. However, in the presence of compression and chirp, the plasma frequency becomes complex. The imaginary part of it will give the growth. This will be what we are seeking for in the future.        

\begin{acknowledgments}
The authors wish to acknowledge the useful discussions with Dr. Lei Shen at SINAP on the plasma physics, Dr. Chao Feng, and many other colleagues in SINAP for the help on the numerical methods. The work is supported by National Natural Science Foundation of China (NSFC), grant No. 11275253 and US DOE, contract DE-FG02-92ER40747.
\end{acknowledgments}

\end{document}